\documentclass[aps,amsmath,amssymb,nofootinbib,superscriptaddress,showpacs,floatfix,prb,twocolumn]{revtex4-1}

\usepackage{latexsym}
\usepackage{graphicx}
\usepackage{times,psfrag,subfigure}
\usepackage{amsmath}
\usepackage{dsfont}
\usepackage{dcolumn}
\usepackage{bm,bbm}       
\usepackage{color}
\usepackage{latexsym,amsmath,amssymb,bm,euscript}
\def\sgn{\mathop{\textrm{sgn}}}

\newcommand{\beq}{\begin{equation}}
\newcommand{\eeq}{\end{equation}}
\newcommand{\beqarray}{\begin{eqnarray}}
\newcommand{\eeqarray}{\end{eqnarray}}

\begin{document}

\allowdisplaybreaks

\title{Stability of  flat-band edge states in  topological superconductors without inversion center}

\date{\today}

\author{Raquel Queiroz}
\email{r.queiroz@fkf.mpg.de}
\affiliation{Max-Planck-Institut f\"ur Festk\"orperforschung,
  Heisenbergstrasse 1, D-70569 Stuttgart, Germany}

\author{Andreas P. Schnyder}
\email{a.schnyder@fkf.mpg.de}
\affiliation{Max-Planck-Institut f\"ur Festk\"orperforschung,
  Heisenbergstrasse 1, D-70569 Stuttgart, Germany}

\begin{abstract}
Nodal superconductors without inversion symmetry exhibit nontrivial topological properties, manifested by topologically protected flat-band edge states. Here we study the effects of edge roughness and strong edge disorder on the flat-band states using large-scale numerical simulations. 
We show that the bulk-edge correspondence remains valid for rough edges and demonstrate that midgap states generically appear at the boundary of nodal noncentrosymmetric superconductors, for almost all edge orientations.
Moderately strong nonmagnetic disorder shifts  some of the edge states away from zero energy, but does not change their total number.
Strong  spin-independent edge disorder, on the other hand, leads to the appearance of new weakly disordered midgap states in the layers adjacent to the  disordered edge, i.e., at the interface between the bulk topological superconductor and the one-dimensional Anderson insulator formed by the strongly disordered edge layers. Furthermore, we show that magnetic impurities, which lift the time-reversal symmetry protection of the flat-band states,
lead to a rapid decrease of the number of edge states with increasing disorder strength. 
\end{abstract}

\date{\today}

\pacs{03.65.vf,74.50.+r, 73.20.Fz, 73.20.-r:}


\maketitle

\section{Introduction}

Topological superconductors have recently attracted considerable theoretical~\cite{schnyderPRB08,kitaev22,hasanKaneRMP,ryuNJP2010,schnyderPRB10,qiZhangRMP11,beenakkerReview,aliceaRepProg} and experimental~\cite{sasakiAndoPRL11,Mourik25052012,heidblumNatPhys12,denXuNanoLett12,mondalPRB12} interest, due to the possibility of realizing 
exotic zero-energy edge states in these systems. Depending on the superconducting pairing symmetry, these edge states
are either chiral or helical Majorana modes, or, in the case of nodal superconductors, form zero-energy flat bands.\cite{Schnyder10,tanakaPRL2010,satoPRB2011,Brydon10,dahlhaus2012,wongLeePRB13,yadaPRB2011,beriPRB10,Schnyder12,matsuura2012}  One particularly interesting
class of topological superconductors are noncentrosymmetric superconductors with strong spin-orbit coupling.\cite{bauerSigristBook,fujimotoJPSJ07}
Many of these compounds, e.g.,  CePt$_3$Si,\cite{bauerPRL04,IzawaPRL05,bonaldeNJP09} CeIrSi$_3$,\cite{Sugitani06,OnukiPRL08} and Li$_2$Pt$_3$B,\cite{yuanPRL06,nishiyama07,badicaJPSJ05,eguchiMaenoPRB13} are reported to have unconventional pairing symmetries with sizable spin-triplet pairing components and line nodes in the superconducting gap. As a result of Rashba-type spin-orbit interactions,
the flat-band edge states in these systems are spin-nondegenerate and exhibit a  helical spin polarization,
where the spin orientation varies as a function of edge momentum.\cite{tanakaNagaosaPRB09,brydonNJP2013,Schnyder13,Hofmann13} 
As a consequence of the nontrivial spin texture, the lowest-order matrix element for spin-independent backscattering among
 the surface states is suppressed.\cite{Schnyder13,Hofmann13} 

The boundary states of clean topological superconductors in a ribbon or slab geometry are well studied theoretically. For example, for a
two-dimensional  ($d_{xy}+p$)-wave superconductor on the square lattice, it was shown that flat-band edge states appear at the (10) and \mbox{(01) edges}, but are absent at the (11) edge (see Fig.~\ref{fig1L}). This can be understood in terms of a bulk-edge  correspondence: 
The topological properties of the  quasiparticle wave functions in the bulk, which are 
characterized by a one-dimensional winding number, directly imply the existence of zero-energy states at the edge.\cite{yadaPRB2011,Schnyder10,Schnyder12,matsuura2012,grafArxiv2012} 
However, in the presence
of edge disorder or for a superconducting dot with a closed boundary,  it is not clear whether the bulk-edge correspondence still applies. Strictly speaking, 
the topological winding number is ill-defined in the absence of translational symmetry, 
since it is given in terms of a momentum-space integral.\cite{footnote00} Nevertheless, 
sufficiently large disks of topological superconductors are expected to show the same edge properties
as topological superconductors in an infinite ribbon geometry.
The study of edge roughness or edge disorder has direct relevance for experiments, since surfaces of unconventional superconductors are often either intrinsically disordered or can  be intentionally disordered via the deposition of impurity atoms. 

In this paper, using the ($d_{xy}+p$)-wave superconductor as a prototypical example, 
we study the edge properties of  disordered nodal topological superconductors with rough irregular boundaries
consisting of both (10) edge and (11) edge parts. By means of large-scale numerical simulations of a two-dimensional
Bogoliubov-de Gennes (BdG) lattice model, we demonstrate  the validity of the 
bulk-edge correspondence for rough edges
, and show that the number of edge states is proportional to the length of the boundary (see Fig.~\ref{fig2L}). 
Secondly, we investigate the effects of strong nonmagnetic edge impurities, which influence both
the edge and the bulk quasiparticle wave functions, leading to a nontrivial coupling between the two. 
We find that strong  edge disorder localizes the states in the edge layer, but leads to the appearance of new weakly disordered ingap states in the second and third inward layers, just below the strongly disordered edge. That is, zero-energy states appear at the interface
between the bulk topological superconductor and the Anderson insulating state of the first layer (see Fig.~\ref{fig4L}).
Weak nonmagnetic impurities, on the other hand, only spread the zero-energy edge states over an energy band of small finite width, 
leaving the total number of edge states unchanged. 
Finally, we also consider magnetic impurities, which lift the 
symmetry protection of the flat-band edge states.
Due to spin-flip scattering, a finite density of impurity spins gives rise to a rapid decrease of the number of edge states  with
increasing disorder strength (see Fig.~\ref{fig3L}).

The outline of the paper is as follows. In Sec.~\ref{sec:NCS} we start by introducing the BdG Hamiltonian of the ($d_{xy}+p$)-wave superconductor
on the square lattice. In  Sec.~\ref{EdgesAnalytics} an analytical expression for the flat-band edge-state wave functions is given and the lowest-order matrix element for scattering among the flat-band edge states is determined.
Using large-scale exact diagonalization, we investigate in Sec.~\ref{Sec:QMdot} the edge properties of  superconducting dots with closed irregular boundaries. The effects of strong magnetic and nonmagnetic edge disorder are studied numerically in Sec.~\ref{sec:dis} by means of 
the recursive Green's function technique. We conclude with a brief discussion in Sec.~\ref{sec:Conclu}. 
Some of the technical details are relegated to two Appendices.

\section{Model definition}
\label{sec:NCS}

We study the stability of  flat-band edge states in noncentrosymmetric superconductors by considering, as a representative example, the ($d_{xy}+p$)-wave superconductor on the square lattice with both spin-singlet and spin-triplet pairing components.\cite{tanakaPRL2010}
In momentum space this topological superconductor is described by a $4 \times 4$ 
BdG Hamiltonian  $\mathcal{H} = \frac{1}{2} \sum_{\bf k}\Phi_{\bf
    k}^{\dagger}H^{\ }_{\bf k}\Phi^{\ }_{\bf k}$, with 
\begin{subequations} \label{modelDef}
\begin{eqnarray}
\label{TBmodel}
H_{\bf k} = \begin{pmatrix}
h_{\bf k}  &
\Delta_{\bf k} \cr
\Delta^{\dag}_{\bf k} &  
- h^{\mathrm{T}}_{- \bf k} 
\end{pmatrix} \label{ham},
\end{eqnarray}
and the four-component Nambu spinor $\Phi_{\bf k} = ( c_{{\bf k} \uparrow},  c_{{\bf k} \downarrow},  c^\dag_{-{\bf
k} \uparrow} , c^\dag_{-{\bf k} \downarrow}  )^{\mathrm{T}}$, where
$c^{\dag}_{{\bf k}\sigma}$ ($c^{\ }_{{\bf k}\sigma}$) creates (annihilates) an electron with
spin $\sigma$ and momentum ${\bf k}$. The normal-state dispersion of the
electrons is given by $h_{\bf k}=\varepsilon_{\bf k} \sigma_0  + \lambda \, {\bf l}_{\bf k} \cdot \bm{\sigma}$, where
$\varepsilon_{\bf k} = t\, ( \cos k_x + \cos k_y) - \mu$,  $\bm{\sigma} = ( \sigma_x , \sigma_y, \sigma_z )^{\textrm{T}}$ is the vector of Pauli matrices, $\sigma_0$ the $2 \times 2$ identity matrix,
and
${\bf l}_{\bf k}$ represents the Rashba-type spin-orbit coupling potential with ${\bf l}_{\bf k} = \hat{\bf x} \sin k_y - \hat{\bf y} \sin k_x$.
Here,  $t$ denotes twice the nearest-neighbor hopping integral,  $\mu$ is the chemical potential, and 
$\lambda$ stands for the spin-orbit coupling strength.
Due to the absence of inversion symmetry, the superconducting order parameter $\Delta_{\bf k}$ contains both even-parity spin-singlet and odd-parity spin-triplet pairing components 
\begin{eqnarray} \label{def:Gap}
\Delta_{\bf k} = f_{\bf k}\left(  \Delta_{\textrm{s}} \sigma_0 + \Delta_{\textrm{t}} {\bf l}_{\bf k} \cdot
\bm{\sigma} \right) ( i \sigma_y) ,
\end{eqnarray}
\end{subequations}
with $f_{\bf k}=\sin k_x \sin k_y $, and where $\Delta_{\textrm{s}}$ and $\Delta_{\textrm{t}}$ denote
the spin-singlet and spin-triplet pairing amplitudes, respectively. 
Unless otherwise specified, we set
$(t, \mu , \lambda, \Delta_\textrm{s}, \Delta_\textrm{t}) = (2.0,2.0,  1.0,0.0,1.0)$ for our numerical calculations.
We have checked that different parameter choices do not qualitatively alter our results, as long as the nodal structure of the superconductor remains the same.

\begin{figure}[t!]
\centering
\includegraphics[clip,angle=0,width=1\columnwidth]{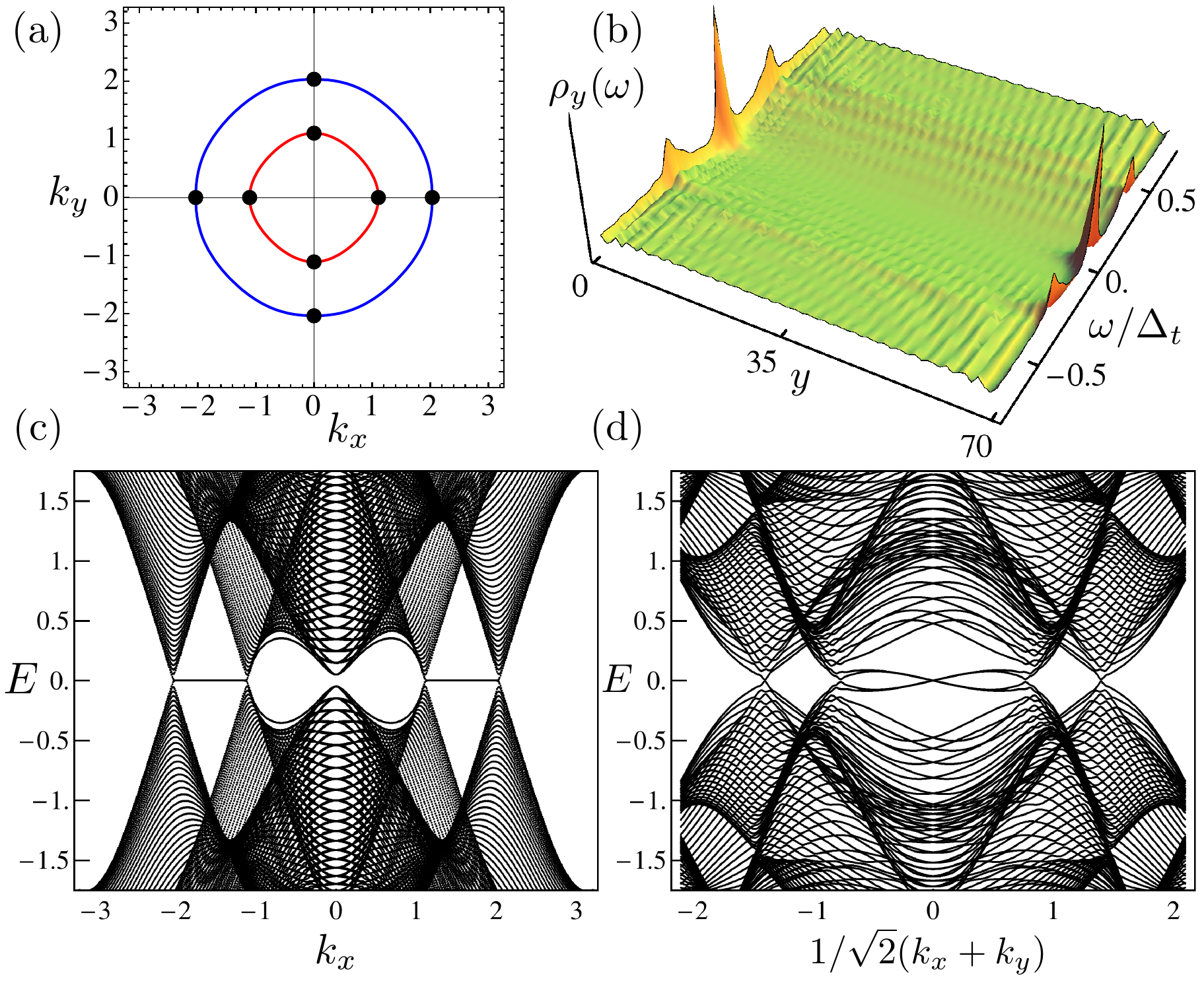}
\caption{\label{fig1L}  
(Color online)  (a) Spin-orbit split Fermi surfaces and nodal points of the superconducting gap $\Delta_{\bf k}$, Eq.~\eqref{modelDef}.
Solid black dots indicate the
location of the eight nodal points, where the gap of the quasiparticle spectrum closes. 
(b) Energy- and layer-resolved density of states $\rho_y ( \omega )$ of the ($d_{xy}+p$)-wave superconductor in a ribbon geometry with (01) edge and width $\mathcal{N}_y=70$ lattice sites. 
(c) and (d): Edge band structure of the ($d_{xy}+p$)-wave superconductor  in a ribbon geometry with
(01) and (11) edge, respectively, as a function of edge momentum.
Zero-energy flat bands appear at the (01) edge connecting the 
projected nodal points of the two helical Fermi surfaces. 
These flat-bands give rise to a divergent density of states at $\omega = 0$, see panel (b).
Note that there are also dispersing edge states, which lead to a feature in the edge density of states
at $\omega\simeq \pm 0.3$.
}  
\end{figure}

The Rashba-type spin-orbit coupling $\lambda {\bf l}_{\bf k}$ splits the normal-state Fermi surface into two helical Fermi surfaces, given by
$\xi^{\pm}_{\bf k}  = \varepsilon_{\bf k} \pm \lambda \left| {\bf l}_{\bf k} \right| = 0$. As a result of the $d_{xy}$-wave form factor $f_{\bf k}$, the 
superconducting order parameter $\Delta_{\bf k}$ on these two helical Fermi surfaces
changes sign, leading  to eight nodal points where the gap of the quasiparticle spectrum vanishes [see Fig.~\ref{fig1L}(a)].
These gap closing-points are located at  
$( \pm k^{\alpha}_0, 0 )$ and $(0, \pm k^{\alpha}_0 )$
in the two-dimensional Brillouin zone,
where  
\begin{align}
k^{\alpha}_0
=
\arccos \left[
\frac{ t ( \mu  - t )  + \alpha \lambda \sqrt{ \lambda^2 +  \mu ( 2 t - \mu )  } }
{  t^2 +  \lambda^2 } 
\right] ,
\end{align} 
with $\alpha \in \left\{ +, - \right\}$. 
The gapless quasiparticle band structure of these Dirac points is protected by a combination of time-reversal and particle-hole symmetry, and their stability is guaranteed by the conservation of a quantized topological invariant [cf.~Eq.~\eqref{eq:Wno}].\cite{beriPRB10,yadaPRB2011,Schnyder10,Schnyder12,matsuura2012}  Particle-hole symmetry $\mathcal{C} = U_C \mathcal{K}$
and time-reversal symmetry $\mathcal{T}= U_T \mathcal{K}$ act on the BdG Hamiltonian~\eqref{modelDef} as 
$U_C H^{\mathrm{T}}_{- \bf k} U^{\dag}_C = -H_{ {\bf k}}$ 
and $U_T  H^{\mathrm{T}}_{- \bf k} U_T^\dag= + H_{{\bf k}}$, respectively,
where  $\mathcal{K}$ is the complex conjugation operator,
$U_C =\sigma_x\otimes\sigma_0$, and $U_T =  \sigma_0 \otimes i\sigma_y$.
Since $\mathcal{C}^2 = +1$ and  $\mathcal{T}^2 = -1$, Hamiltonian~\eqref{modelDef} belongs to
class DIII of the symmetry classification.\cite{altlandPRB97}  
As a result of these symmetries, $H_{ {\bf k}}$ anticommutes with the unitary matrix 
$U_S = i \mathcal{T} \mathcal{C} = - \sigma_x \otimes \sigma_y$, i.e., 
 $U_S ^{\dag} H_{\bf k}  U^{\phantom{\dag}}_S = - H_{\bf k} $. 
Hence, $H_{\bf k} $ takes off-diagonal form in the basis in which $U_S$ is diagonal. That is, we have 
$\widetilde{U}_S = W U_S W^\dag =  \mathrm{diag} ( \sigma_0, - \sigma_0 )$, 
with the transformation matrix
\begin{subequations} \label{eq:OffdiagB}
\begin{eqnarray}
W =\frac{1}{\sqrt{2}}\begin{pmatrix}
\sigma_0&-\sigma_y\\ \sigma_0 & \sigma_y
\end{pmatrix} .
\end{eqnarray}
The transformed BdG Hamiltonian reads
\begin{eqnarray}
&&
\widetilde{H} ( {\bf k} ) = W H_{\bf k} W^\dag=\begin{pmatrix}
0&D ({\bf k} ) \\D^\dag ( {\bf k} ) &0
\end{pmatrix}\label{blockham}, 
\end{eqnarray}
\end{subequations}
where $D ( {\bf k} ) =h_{\bf k} \sigma_0 +\Delta_{\bf k}\sigma_y$.
Due to  the chiral symmetry $U_S ^{\dag} H_{\bf k}  U^{\phantom{\dag}}_S = - H_{\bf k} $, we can choose the zero-energy eigenfunctions of $H_{\bf k}$ to be
simultaneous eigenstates of $U_S$ with a defined chirality eigenvalue $\Gamma = + 1$ or $\Gamma = -1$.\cite{satoPRB2011}

The topological invariant that guarantees the  stability of the eight nodal points of $H_{\bf k}$
can be defined in terms of the winding number $W_{\mathcal{C}}$ of  $ \det D ( {\bf k} )$,
i.e., in terms of the number of revolutions of  $\det D ( {\bf k} ) $ around the origin of the complex
plane as ${\bf k}$ moves along a closed contour $\mathcal{C}$.
An explicit expression of $W_{\mathcal{C}} \in \mathbbm{Z}$ is given by\cite{yadaPRB2011, beriPRB10,Schnyder12,footnote1}
\begin{eqnarray} \label{eq:Wno}
W_{\mathcal{C}}
=
\frac{1}{2 \piÊ}
\oint_{\mathcal{C}} d k_l \, \partial_{k_l} \left\{  \mathrm{arg} \left[ \det D ( {\bf k} ) \right] \right\} ,
\end{eqnarray}
where $\mathcal{C}$ is a one-dimensional  contour encircling one (or several) nodal points in momentum space. 
As a consequence of the bulk-edge correspondence, a nonzero winding number $W_{\mathcal{C}} \ne 0$ signals  the
appearance of zero-energy modes at certain edges. For example, 
for the (01) edge
one can define an edge momentum-dependent winding number $W_{\mathcal{C}} ( k_x )$ by taking 
$\mathcal{C} = \left\{ ( k_x, k_y) \right| \left. -\pi \leq k_y < \pi \right\}$.  
For $k_x$ between the projected nodes of the two helical Fermi surfaces, $W_{\mathcal{C}} ( k_x )$ evaluates to $+1$ or $-1$,
which gives rise to a spin-nondegenerate zero-energy flat band at these edge momenta [Fig.~\ref{fig1L}(c)].
On the (11) edge, however, zero-energy flat bands are absent [Fig.~\ref{fig1L}(d)].\cite{tanakaPRL2010,satoPRB2011,yadaPRB2011}  

In Sec.~\ref{Sec:QMdot}, we will numerically compute the edge modes of the  ($d_{xy}+p$)-wave superconductor for 
different edges. We will  demonstrate that zero-energy flat-band states generically appear at the boundary for arbitrary edge orientations, except for the (11) edge. But before doing so, we first derive in the following section the lowest-order matrix elements for impurity scattering among the flat-band edge states
using an explicit expression for the edge-state wave functions.
 
 \section{Flat-band edge-state wave function}
\label{EdgesAnalytics}

To derive the zero-energy edge-state wave functions, let us consider Hamiltonian~\eqref{modelDef} 
on the semi-infinite plane $y>0$, with the (01) edge located at $y = 0$.
The ansatz for the nondegenerate edge-state wave function is taken to be
$\Psi_{k_x}  = \sum_{\alpha, \beta} C^{\alpha}_{\beta}  \psi^{\alpha}_{\beta} e^{\kappa^{\alpha}_{\beta} y}e^{i k_x x}$, which decays exponentially into the bulk
with inverse decay lengths $\mathrm{Re} [ \kappa^{\alpha}_{\beta} ] < 0$. 
In the off-diagonal 
basis, Eq.~\eqref{eq:OffdiagB},  the wave function 
  $\Psi_{k_x}  = \left( \chi_{k_x} , \eta_{k_x}  \right)^{\mathrm{T}}$
can be split into a part  $( \chi_{k_x}  , 0)^{\textrm{T}}$ with positive chirality $\Gamma = +1$ 
and a part $(0, \eta_{k_x}  )^{\textrm{T}}$ with  negative chirality $\Gamma = -1$.
Since all the eigenstates of $H_{\bf k}$ can be chosen to have definite chirality, it follows that $\eta_{k_x}  = 0$ whenever $\chi_{k_x}  \ne 0$ and vice versa.\cite{satoPRB2011,tanakaPRL2010,Schnyder10} 
In addition, we observe that for every edge-state wave function 
\begin{subequations} \label{FlatBandExpr1}
\begin{eqnarray}  \label{FlatBandExpr1A}
\Psi^+_{k_x} = \left( \chi_{k_x}   , 0 \right)^{\textrm{T}} 
\end{eqnarray}
with edge momentum $k_x$ and $\Gamma = +1$ there is a time-reversed partner 
\begin{eqnarray} \label{FlatBandExpr1B}
\Psi^-_{k_x}  = \left( 0, i \sigma_y \chi^{\ast}_{-k_x} \right)^{\textrm{T}}  
\end{eqnarray}
with edge momentum $-k_x$ and $\Gamma = -1 $. 
Using quasiclassical scattering theory, it is shown in Appendix~\ref{app000} that the zero-energy edge-state wavefunction 
$\Psi^+_{k_x} = ( \chi_{k_x} , 0)^{\textrm{T}} $ for $- k^-_{\textrm{F}} < k_x < - k^+_{\textrm{F}}$ and
$\Delta_\textrm{t} > \Delta_\textrm{s}$
can be explicitly written as\cite{satoPRB2011,Schnyder12} 
\begin{eqnarray}  \label{eqChiExplicit}
\chi_{k_x}  = 
\begin{pmatrix}
- 2 k^+_{\textrm{F}} k^-_{\perp} e^{\kappa^+_1 y}  + b_1 e^{\kappa^-_2 y} + b_1^{\ast} e^{\kappa^-_1 y}   \cr
 2 i a_1 k^-_\perp e^{\kappa^+_1 y}    - a^{\ }_2 i e^{\kappa_1^- y}  - a^{\ast}_2 i e^{\kappa_2^- y} \cr 
\end{pmatrix} 
e^{i k_x x} ,
\end{eqnarray}
\end{subequations}
where $k^-_{\perp} = \sqrt{  ( k_{\textrm{F}}^-)^2  - k_x^2}$ and $k^{\pm}_{\textrm{F}}$ denote the Fermi momenta of the two helical Fermi surfaces.
The coefficients $a_1$, $a_2$, and $b_1$  depend on the edge momentum $k_x$ and are defined below Eq.~\eqref{EdgwWaveFunA} in Appendix~\ref{AppQuasiClassics}.
The inverse decay length $\kappa_1^+$ is purely real, whereas $\kappa_1^-$ and $\kappa_2^-$ are complex conjugate partners,
see Eq.~\eqref{defKappa}. 
An expression similar to Eq.~\eqref{FlatBandExpr1} can be derived for the zero-energy edge states on the opposite edge, i.e., for 
Hamiltonian~\eqref{modelDef}  on $y<0$, which supports
zero-energy flat band states with opposite chirality as compared to Eq.~\eqref{FlatBandExpr1}.

 \subsection{Spin polarization of  flat-band edge states}

Surface states of noncentrosymmetric superconductors exhibit a helical spin texture, where the 
spin orientation of the surface states is correlated with their momentum.\cite{vorontsovPRL08,luYip10,brydonNJP2013,Schnyder13,Hofmann13} 
For the (01) edge of the ($d_{xy}+p$)-wave superconductor one finds that the flat-band states are
strongly polarized in the $yz$-spin plane, 
but have a vanishing spin component along the $x$ axis.\cite{brydonNJP2013,Schnyder13,Hofmann13} 
Using Eq.~\eqref{FlatBandExpr1}, it can be explicitly verified that the expectation value of the spin operator
\begin{eqnarray}
S^{\mu} 
=
\begin{pmatrix}
\sigma^{\mu} & 0 \cr
0 & - \left[ \sigma^{\mu} \right]^{\ast} \cr
\end{pmatrix},
\qquad
\mu \in \left\{ x, y, z \right\} ,
\end{eqnarray}
with respect to the surface-state wave functions $\Psi^{\pm}_{k_x}$, Eq.~\eqref{FlatBandExpr1}, 
has the following properties (cf.\ Appendix~\ref{app000})
\begin{eqnarray}
&&
\left\langle \Psi^{\pm}_{k_x}  \right|   \widetilde{S}^{x} \left| \Psi^{\pm}_{k_x}  \right\rangle  =0,
\quad
 \left\langle \Psi^\pm_{k_x}  \right|   \widetilde{S}^{y,z} \left| \Psi^{\pm}_{k_x}  \right\rangle  \ne 0, 
\end{eqnarray}
for all $k_x$ with $  k^+_{\textrm{F}} <  | k_x | <  k^-_{\textrm{F}}$. Here, 
$ \widetilde{S}$ denotes the spin operator in the off-diagonal basis, i.e., 
$ \widetilde{S}^{\mu}= W S^{\mu} W^{\dag}$.
Moreover, one finds that the $y$ component of the spin expectation value is much larger than the $z$ component
and  that the sign of the $y$-spin polarization correlates with the chirality $\Gamma$ of the flat-band edge state, such that
$
\sgn [  \langle \Psi^{\pm}_{k_x}  |   \widetilde{S}^{y}  | \Psi^{\pm}_{k_x}  \rangle ] = \mp 1 $.  
Finally, we note that the two flat-band edge states $\Psi^{+}_{-k_x}$ and $\Psi^{-}_{k_x}$, which have opposite edge momenta,
have opposite spin polarization. That is,
\begin{eqnarray}
 \left\langle \Psi^{+}_{ - k_x}  \right|   \widetilde{S}^{y,z} \left| \Psi^{+}_{- k_x}  \right\rangle  
 =
- \left\langle \Psi^{-}_{k_x}  \right|   \widetilde{S}^{y,z} \left| \Psi^{-}_{k_x}  \right\rangle ,
\end{eqnarray}
for all $k_x$ with $  k^+_{\textrm{F}} <    k_x <  k^-_{\textrm{F}}$,
which is  consistent with time-reversal symmetry.

\subsection{Impurity scattering among flat-band edge states}
\label{impScatFlat}

In order to calculate the matrix elements for impurity scattering among flat-band edge states,
 we  consider uncorrelated edge disorder 
described by
\begin{eqnarray} \label{EqImpurePo}
H^{\beta}_{\textrm{imp}} 
=
\sum_{ {\bf k}, {\bf q}}
\Phi^{\dag}_{\bf k} V^{\beta}_{q_x} \Phi{\phantom{\dag}}_{{\bf k}+ \hat{{\bf e}}_x q_x} ,
\end{eqnarray}
where 
$
V^{\beta}_{q_x} 
=
(1/ \mathcal{N} ) \sum_j v ( x_j ) \mathcal{S}^{\beta} e^{-i q_x  x_j } 
$
denotes the Fourier transform of the impurity potentials $v ( x_j ) \mathcal{S}^{\beta}$ at the edge sites $x_j$ with
strengths $v ( x_j )$. Here, $\mathcal{N}$ stands for the number of lattice sites and $V^{\beta=0}$ corresponds to nonmagnetic impurities
with $\mathcal{S}^0 = \sigma_z \otimes \sigma_0$, while $V^{\beta = x,y,z }$ represents magnetic exchange scattering
with  $\mathcal{S}^{x,y,z} = S^{x,y,z}$. 
First, we observe that impurity scattering processes connecting flat-band edge states to bulk nodal quasiparticles are
strongly suppressed, since the bulk density of states vanishes linearly as $\omega \to 0$ [see Fig.~\ref{fig1L}(b)].
A rough estimate for the effects of impurity scattering among the zero-energy edges states can be obtained
from the matrix elements
of the  impurity potential $V_{q_x}^{\beta}$
between two flat-band  edge-state wavefunctions.
Because the edge spectrum of the ($d_{xy}+p$)-wave superconductor has in general two flat bands with opposite chirality [see Fig.~\ref{fig1L}(c)], 
it is useful
to distinguish between ``intraband" scattering between states with the same chirality and ``interband" scattering between states with opposite chirality.
From 
\begin{eqnarray}
V^0_{q_x} U_S + U_S V^0_{q_x}  =0
\end{eqnarray}
and
\begin{eqnarray}
 \big\langle \Psi^{\pm}_{  k^{\phantom{\prime} }_x} \big| \widetilde{V}^0_{k^{\prime}_x - k^{\phantom{\prime}}_x } \big| \Psi^{\pm}_{ k^{\prime }_x}  \big\rangle   = 0 ,
\quad \textrm{for all $k^{\phantom{\prime}}_x$, $k^{\prime}_x$} ,
\end{eqnarray}
where $\widetilde{V}^0_{q_x} = W V^0_{q_x} W^{\dag}$, 
it follows that  edge flat bands are protected against nonmagnetic intraband scattering by chiral symmetry. In other words, since nonmagnetic onsite disorder preserves the total chirality number of the superconductor, it can remove edge states only in pairs of opposite chirality.
Magnetic impurities, on the other hand, break chiral symmetry, i.e., $V_{q_x}^{x,y,z} U_S + U_S V_{q_x}^{x,y,z} \ne 0$,
 and therefore allow for strong intraband scattering. 

For the case of impurity scattering between edge flat bands with opposite chirality we find by use of Eqs.~\eqref{FlatBandExpr1A} and \eqref{FlatBandExpr1B} that
time-reversal invariance forbids nonmagnetic backscattering between the 
time-reversed partners $\Psi^-_{k_x}$ and $\Psi^+_{-k_x}$. That is, 
\begin{eqnarray} \label{NoBackscattering}
\big\langle \Psi^+_{- k_x } \big|  \widetilde{V}^0_{2 k_x}  \big| \Psi^-_{k_x} \rangle
= 0,
\end{eqnarray}
for all $k_x$ with $k^+_{\textrm{F}} < k_x < k^-_{\textrm{F}}$. Moreover, for  two flat-band edge states 
with nearly opposite momenta $k_x$ and $-k^{\prime}_x$ (i.e., $0 < | k_x - k^{\prime}_x | <  k^{-}_{\textrm{F}} - k^{+}_{\textrm{F}}  $), one finds that 
the corresponding matrix element 
$ \big\langle \Psi^+_{- k^{\prime}_x } \big|  \widetilde{V}^0_{ k^{\phantom{\prime}}_x + k^{\prime}_x}  \big| \Psi^-_{k^{\phantom{\prime}}_x} \rangle$
  is nonzero but small, due to the mismatch between the almost opposite spin polarizations of the two edge states. In the presence of magnetic impurities, however, spin-flip scattering is allowed, and hence
 scattering between  states with opposite spin polarizations is possible.

The above considerations suggest that moderately strong nonmagnetic disorder, with disorder strength $\gamma_{\textrm{imp}}$ smaller
or of the same order as 
 the superconducting gaps $\left| \Delta_{\pm} \right| = \left| \Delta_\textrm{s} \pm \Delta_\textrm{t} \right|$,
has only weak effects on the flat-band edge states.
Magnetic impurities, however, which lift the symmetry protection of the flat-band states, 
are expected to strongly reduce the number of
edge states. To test these expectations, we perform in the following two sections numerical simulations
of the ($d_{xy}+p$)-wave  superconductor in the presence of different types of edge disorder.

\section{Edge states at irregularly shaped boundaries}
\label{Sec:QMdot}

\begin{figure}[t!]
\centering
\includegraphics[clip,angle=0,width=1\columnwidth]{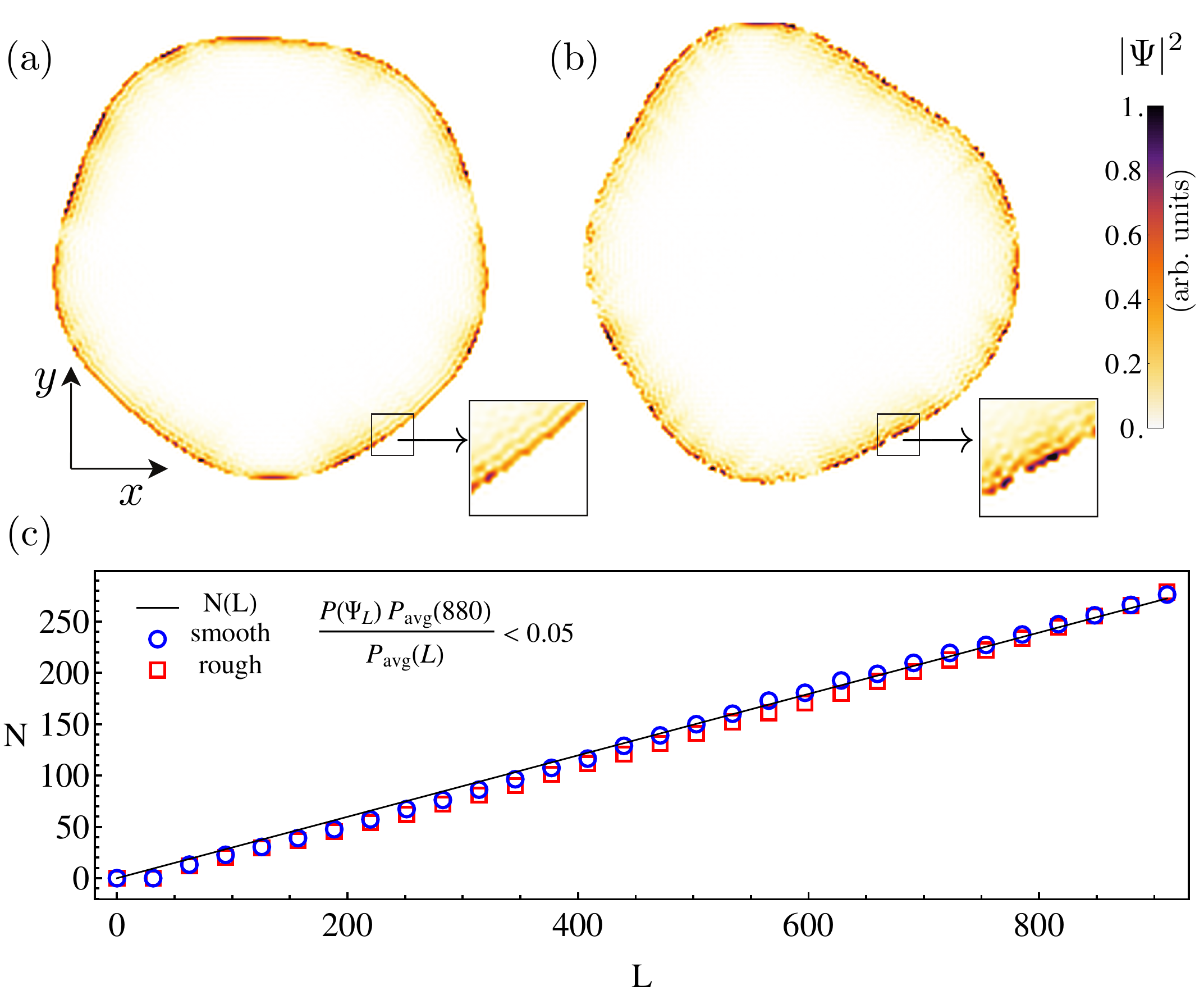}
\caption{\label{fig2L} (Color online) 
Density plot of the edge-state wave function amplitudes $|\Psi_L|^2$ in a  
 ($d_{xy}+p$)-wave superconducting dot
with (a) smooth and (b) rough boundaries. 
Edge states are present both at an irregular but smooth boundary (a) and at a boundary with short-range disorder (b). 
Panel (c) shows the average number of edge states for an ensemble of randomly shaped superconducting dots with
smooth (blue circles) and rough (red squares) boundaries as a function of circumference $L$ of the dot.\cite{footnote2}
Here, the edge states  are separated from the bulk states according to criterion~\eqref{ParticipCrit} and by
additionally requiring that the energy of the states is smaller than $0.1 | \Delta_{\pm} |$ in absolute value.
The solid black line represents 
the analytical approximation
given by Eq.~\eqref{EqNoEdgeSt}. }
\end{figure}

In order to compute the edge-state wave functions of 
an irregularly shaped ($d_{xy}+p$)-wave superconduting dot 
with smooth or rough edges,
we Fourier transform Hamiltonian~\eqref{modelDef} to real space and diagonalize it
using standard  eigenvalue algorithms.\cite{ARPACK}
The shape  of the superconducting dot is defined in terms of a direction-dependent radius~\cite{Wimmer10}
\begin{eqnarray} \label{DotShapes}
R(\theta)=\sum_{i=1}^5w_i\sin(i\theta-\phi_i) ,
\end{eqnarray}
with the parameters $w_i$ and $\phi_i$ and the angle of direction $\theta$.  
With this definition, the  dots can be constructed by cutting
the shapes given by Eq.~\eqref{DotShapes} out of a  square lattice grid.
This results in superconducting dots with smooth edges, whose orientation is locally well defined [Fig.~\ref{fig2L}(a)]. 
Bulk- and edge-state wave functions can be distinguished in terms of the participation ratio $P(\Psi_L)$ of a given
eigenstate $\Psi_L ( {\bf r}_i )$ of a dot with circumference $L$, i.e.,\cite{Bell70} 
\begin{eqnarray}
P(\Psi_L)
=
\frac{ \left( \sum_i \left| \Psi_L ( {\bf r}_i ) \right|^2 \right)^2 }
{ \mathcal{N} \sum_i \left| \Psi_L ( {\bf r}_i ) \right|^4 },
\end{eqnarray}
where $i$ runs over all the sites ${\bf r}_i$ in the dot and $\mathcal{N}$ is the total number of sites. 
 The participation ratio $P(\Psi_L)$ represents the number of lattice sites occupied 
 by the Bogoliubov quasiparticle wave function $\Psi_L$ compared to the total number of sites~$\mathcal{N}$.
 Hence, for extended bulk states $P(\Psi_L)  \simeq 1$,  whereas  for localized edge states 
 $P(\Psi_L) \ll 1$. We find that for sufficiently large dots, a good characterization of the edge-state wave functions 
 is given by 
 \begin{eqnarray} \label{ParticipCrit}
P ( \Psi_L ) \frac{  P_{\textrm{avg}} ( 880) }{ P_{\textrm{avg}} (L) } < 0.05, 
 \end{eqnarray}
 where
$
 P_{\textrm{avg}} (L) = \frac{1}{n} \sum_{\iota=1}^n \frac{1}{ \left| \mathcal{W}_\iota \right| }  \sum_{\Psi^{\iota}_L \in \mathcal{W}_\iota  }  P ( \Psi^{ \iota}_L  ) 
$
is the average participation ratio of all the low-energy wave functions $\Psi^\iota_L$ of an ensemble of randomly shaped
superconducting dots of circumference $L$. Here, $n$ denotes the size of the
statistical ensemble and $\mathcal{W}_\iota$ is the set of the first $\sim   L /2$ lowest positive energy wave functions calculated numerically for each sample.\citep{ARPACK}
Since edge disorder leads to a small $L$-dependent decrease of the participation ratio $P(\Psi_L)$ of all the wave functions $\Psi_L$, 
 we have included in Eq.~\eqref{ParticipCrit} the renormalization factor  $P_{\textrm{avg}} ( 880) / P_{\textrm{avg}} (L)$, where
$L=880$ is the circumference of the largest dots.

\paragraph{Smooth edges.}

We first study irregularly shaped dots with smooth boundaries, where the edge orientation is locally well defined [inset of Fig.~\ref{fig2L}(a)]. 
These boundaries consist of both (01)-edge and (11)-edge type parts, leading to long-range correlated disorder.
As exemplified in Fig.~\ref{fig2L}(a), we find that ingap states appear at almost all boundaries of the dot.
That is, the behavior characteristic of the (01)-edge [Fig.~\ref{fig1L}(c)] is generic and qualitatively independent of the edge orientation.
Hence, the number of edge states is expected to scale linearly with the circumference $L$ of the superconducting dot. 
Within a simplified continuum theory, one can show that the density of edge states per unit length for a smooth edge is
approximately given by (cf.~Appendix~\ref{app000})
\begin{eqnarray} \label{NoEdgePerLength}
\frac{ d N}{d l} 
=
 \left|  k_0^-  \sin \varphi -  k_0^+ \cos \varphi \right|
- \left| k_0^- \cos \varphi - k_0^+ \sin \varphi  \right|  , \quad \;
\end{eqnarray}
where $0 \leq \varphi < \pi/4$ is the angle between the local edge orientation and the nearest (01) or (10) direction.
Integrating Eq.~\eqref{NoEdgePerLength} along the circumference, we find that the total number of edge states
for a circular dot is given by 
\begin{eqnarray}  \label{EqNoEdgeSt}
 N(L)
 &=&
 \int_0^L  \frac{dN}{d l} dl
 \\
 &=&
 \frac{8 L }{\pi } 
\left[ \frac{k_0^-   + k_0^+}{\sqrt{2}}  +
\frac{k_0^- -k_0^+ }{2}    -\sqrt{(k_0^- )^2+ ( k_0^+ )^2} \right] .
\nonumber
\end{eqnarray}
 As it turns out, Eq.~\eqref{EqNoEdgeSt} is
  a good approximation for the number of edge states of
an irregularly shaped dot. 
This is revealed in Fig.~\ref{fig2L}(c), which shows the average number of edge states as a function of $L$ for an ensemble of randomly shaped superconducting dots with smooth boundaries (blue circles)\cite{footnote2} together with the analytical result, Eq.~\eqref{EqNoEdgeSt}.
The numerical data and the analytical  curve are in good agreement except for dots with small circumferences, 
with $L < 50$, where finite-size effects become important. 

\paragraph{Rough edges.}
\label{secRough}

Second, we consider rough boundaries with edge disorder on the lattice scale. In order to introduce short-range edge disorder, we start from
the smooth edges, Eq.~\eqref{DotShapes}, and randomly extract edge sites  with probability $p_{\textrm{rm}}=0.01$, while moving around the edge of the dot once.\cite{Wimmer10} This ``etching" process is repeated twenty times, which leads to an irregular boundary
with both long-range and short-range correlated disorder [inset of Fig.~\ref{fig2L}(b)].
The edge-state wave function amplitudes for a  superconducting dot with rough edges is plotted in Fig.~\ref{fig2L}(b). As in the case of smooth edges, we find that edge states appear  at almost all boundaries. That is, short-range edge disorder does not change the total number of edge states, but only shifts some of the edge states away from zero energy. 
This is further evidenced in Fig~\ref{fig2L}(c) (red squares), which shows that the average number of ingap states at a randomly shaped boundary with short-range disorder scales linearly in $L$, and is in good agreement with Eq.~\eqref{EqNoEdgeSt}.

In conclusion,  our numerical simulations of ($d_{xy}+p$)-wave superconducting dots with short-range and long-range edge disorder demonstrate 
that the bulk-edge correspondence remains valid even in the absence of translation symmetry. Ingap states generically
appear at the boundary of  these  superconductors, for almost all edge orientations. 
Due to their topological origin (cf. Sec.~\ref{impScatFlat}), the edge states are robust against nonmagnetic scattering from both short-range and long-range correlated edge disorder.

\section{Strong edge disorder}
\label{sec:dis}

\begin{figure*}[t!]
\centering
\includegraphics[clip,angle=0,width=2.1\columnwidth]{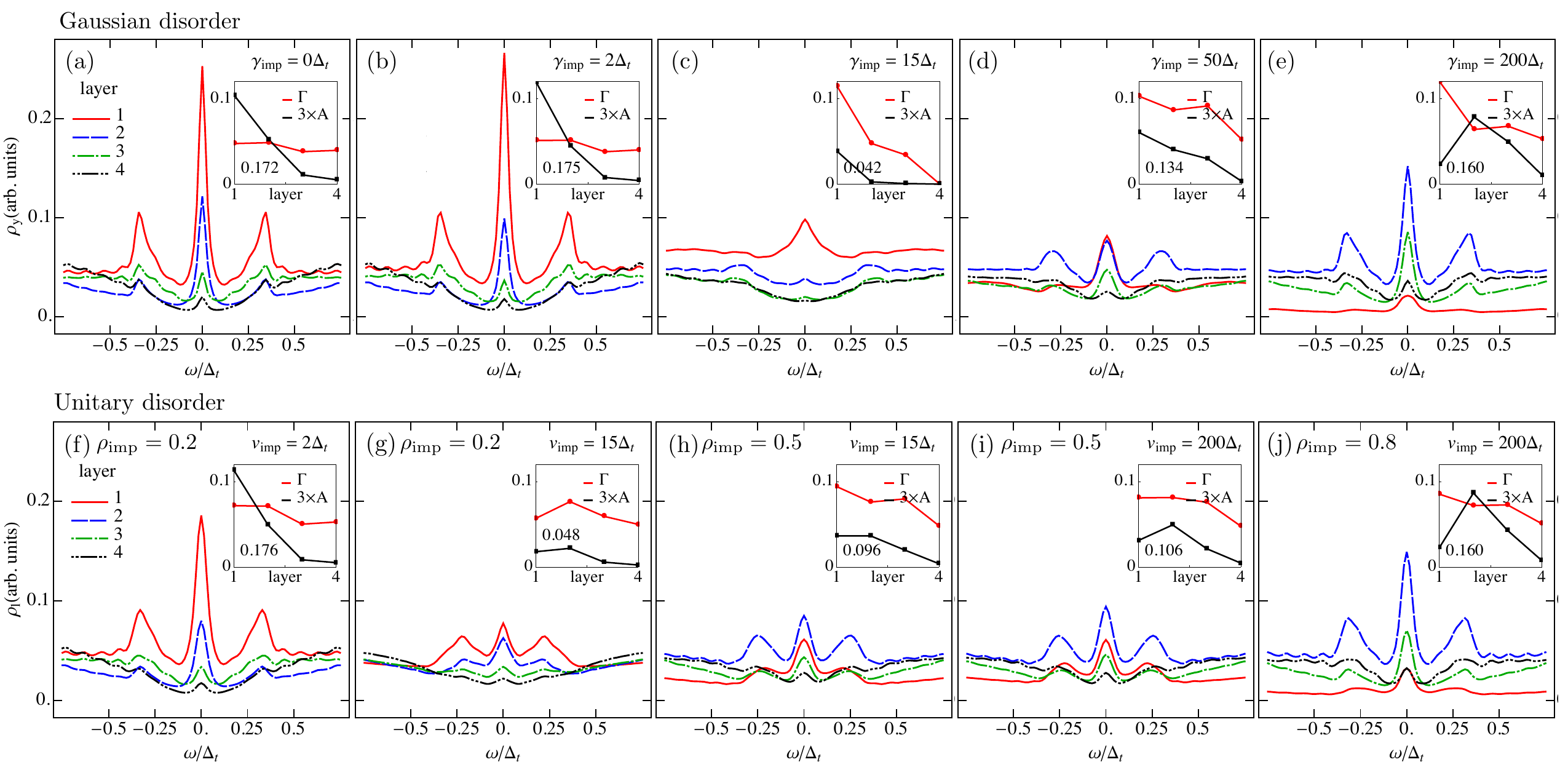}
\caption{\label{fig4L}  
(Color online) 
Layer-resolved local density of states $\rho_y ( \omega )$ plotted for the first four outermost layers of a ($d_{xy}+p$)-wave superconducting ribbon
with (01) edges in the presence of  (a)-(e) ``Gaussian"  edge disorder and (f)-(j) ``unitary" edge disorder (for details see text). 
The insets show the width $\Gamma$ and the area $A$ of the Lorentzian peaks at $\omega = 0$ as a function of layer index $y$.
The number in the lower left corner of the insets indicates the total area of the zero-bias peak as obtained by summing 
 $A$ over the first four layers.
 In the clean case, $v_{\textrm{imp}}=0$, the edge states penetrate only about two layers into the bulk [panel (a)]. For strong disorder
the outermost layer shows signatures of localization, while new weakly disordered states appear in the second and third inward layers
[panels (e) and (j)].
}  
\end{figure*}

\begin{figure*}[t!]
\centering
\includegraphics[clip,angle=0,width=2.1\columnwidth]{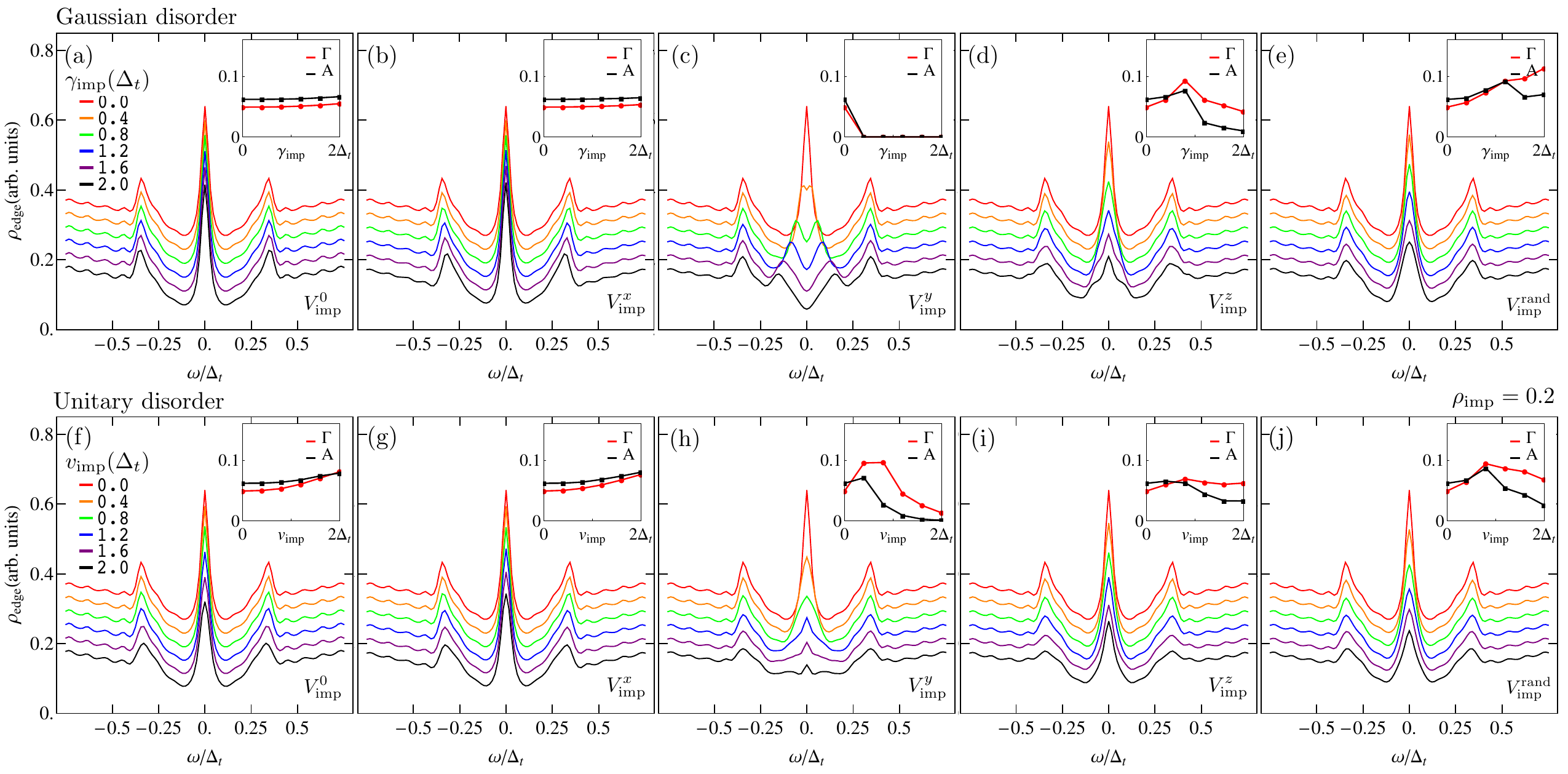}
\caption{\label{fig3L}  
(Color online) 
Local density of states summed over the four outermost layers, $\rho_{\textrm{edge}} (\omega) = \frac{1}{4} \sum_{y=1}^4 \rho_y ( \omega)$, of a  ($d_{xy}+p$)-wave superconducting ribbon with (01) edges in the presence of nonmagnetic [panels (a) and (f)] and magnetic impurities [panels (b)-(e) and (g)-(j)]  in the two outermost layers. Two different disorder distributions are considered: (a)-(e) ``Gaussian" disorder and (f)-(j) ``unitary" disorder with $\rho_{\textrm{imp}}=0.2$ (for details see text).
Individual traces are vertically offset by $0.02$ from one another for clarity. 
 The insets show the width $\Gamma$ and the area $A$ of the Lorentzian peaks at $\omega=0$ as a function of disorder strengths 
 $\gamma_{\textrm{imp}}$ and $v_{\textrm{imp}}$, respectively. 
}  
\end{figure*}

Let us now investigate in detail the effects of strong edge disorder, which affects both edge and bulk states,
leading to a nontrivial interaction between the two. 
 In order to access larger system sizes than in Sec.~\ref{Sec:QMdot}, we employ here recursive Green's function techniques\cite{PotterLee11, LeeFisher81} to calculate the lattice Green's function $G ( \omega ; {\bf r} )$
 of a disordered ($d_{xy}+p$)-wave superconducting ribbon (see Appendix~\ref{appA}). 
 From the Green's function $G ( \omega ; {\bf r} )$ the local density of states in the $y$-th layer is obtained via
\begin{eqnarray} \label{LDOSrgf}
\rho_y ( \omega )
=
- \frac{1}{4 \pi } \frac{ 1}{\mathcal{N}_x} \sum_{x } 
\textrm{Im} \left[
\mathrm{Tr}
\left\{
G ( \omega; x , y) \right\} \right] ,
\end{eqnarray}
where $\mathcal{N}_x$ denotes the length of the superconducting ribbon. 
In the following, we have considered samples of width $\mathcal{N}_y = 70$ sites and length $\mathcal{N}_x = 600$ sites. 
Quenched edge disorder is implemented by adding
 random on-site potentials $V^{\beta}_{x_j} = v ( x_j) \mathcal{S}^{\beta}$ in the two outermost layers
of the superconducting ribbon. We consider two different types of disorder distributions:\cite{SchnyderThesis, Atkinson00,chamonMudry01} (i) scatterers at each lattice site with local potentials $v (x_j)$ 
drawn from a box distribution $p\left[ v (x_j) \right] = 1 / \gamma_{\textrm{imp}}$ for 
$v(x_j) \in \left[ - \gamma_{\textrm{imp}} /2 , + \gamma_{\textrm{imp}} / 2 \right]$ (referred to as ``Gaussian" type disorder),\cite{Atkinson00}
and (ii) a dilute density $\rho_{\textrm{imp}}$ of strong scatterers with
constant potential strength $v ( x_j) \equiv v_{\textrm{imp}} \gtrsim | \Delta_{\pm} |$ (referred to as ``unitary" type disorder). 
In case (i) the strength of the disorder is controlled by the width $\gamma_{\textrm{imp}}$ of the distribution, whereas
in case (ii) it can be adjusted in terms of both the impurity density $\rho_{\textrm{imp}}$ and the potential strength $v_{\textrm{imp}}$.
Since fluctuations between different disorder realizations are minor, we present in Figs.~\ref{fig4L} and~\ref{fig3L}  spectra for a specific disorder realization, without averaging over disorder configurations.

\subsection{Nonmagnetic impurities}

\setcounter{paragraph}{0}

We start by discussing the effects of nonmagnetic impurities with potentials $V^{0}_{x_j} = v ( x_j) \mathcal{S}^{0}$.
In Fig.~\ref{fig4L} is shown the local density of states $\rho_y ( \omega)$ for the first four outermost layers 
of a ($d_{xy}+p$)-wave superconducting ribbon with nonmagnetic disorder of different strengths.
The case of ``Gaussian" type disorder is plotted in panels (a)-(e), whereas the effects of ``unitary" type disorder
are presented in panels (f)-(j). In order to estimate the number of ingap edge states in the system, we have
fitted a Lorentzian function to the zero-bias peaks in Fig.~\ref{fig4L}.
The peak width $\Gamma$ and the peak area $A$ provide a measure for
the number of edge states and their spread in energy, respectively (insets in Fig.~\ref{fig4L}).
In agreement with the analytical arguments given in Sec.~\ref{impScatFlat}, we find that weak and even
moderately strong disorder, with $\gamma_{\textrm{imp}}$ (or $\rho_{\textrm{imp}} v_{\textrm{imp}} $)  
of the same order as the superconducting gaps $\left| \Delta_{\pm} \right|$, has very little effect on the
edge states: Gaussian disorder gives rise to a slightly faster decay of the
edge states into the bulk [Fig.~\ref{fig4L}(b)], whereas unitary disorder 
somewhat increases  the energy spread of the ingap states [Fig.~\ref{fig4L}(f)]. The total number of edge states,
however, is unaffected by moderately strong disorder [compare insets in Figs.~\ref{fig4L}(a),~\ref{fig4L}(b), and~\ref{fig4L}(f)].

For strong edge disorder with $\gamma_{\textrm{imp}} \gg \left| \Delta_{\pm} \right|$ (or $\rho_{\textrm{imp}} v_{\textrm{imp}} \gg \left| \Delta_{\pm} \right|$), on the other hand, the states in the outermost layer become strongly localized.  But remarkably, new weakly disordered edge states
appear at the second and third inward layers [Fig.~\ref{fig4L}(e) and~\ref{fig4L}(j)]. In other words, due to the bulk-boundary
correspondence, zero-energy states
emerge at the interface between the bulk topological superconductor and the Anderson insulator formed by the outermost layer.
This behavior is reminiscent of topological-insulator surface states perturbed by strong disorder.\cite{schubertPRB12,RingelStern}

\subsection{Magnetic impurities}

Magnetic impurities $V^{x,y,z}_{x_j} = v ( x_j) \mathcal{S}^{x,y,z}$ break time-reversal symmetry, thereby lifting the symmetry protection of the edge states. 
In Fig.~\ref{fig3L} we present the edge density of states $\rho_{\textrm{edge}}$, defined
as the sum of  $\rho_y ( \omega)$ over the four outermost layers, of a  ($d_{xy}+p$)-wave superconducting ribbon with (01) edges in the presence of impurity spins polarized along the $x$,  $y$, and $z$ axes [panels (b)-(d) and (g)-(i)] and randomly oriented magnetic disorder [panels (e) and (j)]. For comparison, Figs.~\ref{fig3L}(a) and~\ref{fig3L}(f) show the edge density of states for nonmagnetic scalar impurities.  As before, we consider both ``Gaussian" type disorder [Figs.~\ref{fig3L}(a)-(e)] and ``unitary" type disorder [Figs.~\ref{fig3L}(f)-(j)].
Since the flat-band edge states are polarized within the $yz$ spin-plane (cf.~Sec.~\ref{EdgesAnalytics}), impurity spins polarized along the $y$ and $z$ axes couple strongly to the flat bands, whereas scalar impurities and $x$ spin polarized impurities leave the edge states almost unaffected as long as $\gamma_{\textrm{imp}}$ ($\rho_{\textrm{imp}} v_{\textrm{imp}}$) is not much larger than $\left| \Delta_{\pm} \right|$.
As shown in Figs.~\ref{fig3L}(c) and~\ref{fig3L}(h), $y$ spin polarized impurities are particularly harmful to the flat-band edge states, even for relatively small 
disorder strengths of $\gamma_{\textrm{imp}} \simeq 0.8  \left| \Delta_{\pm} \right|$ (or $v_{\textrm{imp}} \simeq 0.8  \left| \Delta_{\pm} \right|$ 
for the ``unitary" type disorder).

\section{Summary and Conclusions}
\label{sec:Conclu}

In summary, we have shown that flat-band edge states in noncentrosymmetric superconductors are robust against weak and moderately strong nonmagnetic edge disorder, as long as the disorder strength is not much larger than the superconducting gaps.
Using analytical considerations, we have found that spin-independent scattering among the flat-band edge states is suppressed due
to the definite chirality of the edge-state wave functions and their helical spin texture (Sec.~\ref{impScatFlat}). 
By means of extensive numerical simulations, we have demonstrated that
moderately strong spin-independent disorder spreads the zero-energy edge states over a small band in energy, but does not alter the total number of edge states [Figs.~\ref{fig3L}(a) and~\ref{fig3L}(f)]. However, in the presence of strong edge disorder, with disorder strength much larger than the superconducting gaps, the wave functions in the outermost layer localize, but new weakly disordered ingap states appear in the second and third inward layers [Figs.~\ref{fig4L}(e) and \ref{fig4L}(j)]. 
We have investigated the edge orientation dependence of the edge state density by numerically simulating superconducting dots with both smooth and rough boundaries. Edge states  appear for almost all edge orientations, even in the absence
of translation symmetry along the boundary [Fig.~\ref{fig2L}]. This demonstrates 
 that translation symmetry  
 is not crucial for the protection of the edge states. Time-reversal and particle-hole symmetry, on the other hand, play a key role for the stability of the flat-band states. Consequently,  we have found that
 magnetic impurities, which break time-reversal symmetry, substantially decrease the number of edge states even for 
 small impurity densities [Figs.~\ref{fig3L}(c) and~\ref{fig3L}(h)].
 
Nondegenerate flat-band edge states are expected to appear in any nodal topological superconductor with strong Rashba type spin-orbit coupling, such as, e.g., CePt$_3$Si or Li$_2$Pt$_3$B. These boundary states can in principle be observed using scanning tunneling microscopy or angle-resolved photoemission spectroscopy. The signature of the flat-band edge states on transport in various heterostructures involving topological superconductors remains as a direction for future research, as well as the study of interaction effects
among the flat-band edge states.\cite{LiWuNJP13}

\acknowledgments
The authors thank P.~Brydon, C.-K.~Chiu, A.~Damascelli, J.~Hofmann, 
P.~Ostrovsky, S.~Ryu,  C.~Timm, and P.~Wahl for useful discussions. 

\appendix

\section{Derivation of zero-energy edge-state wave function}
\label{app000}
\label{AppQuasiClassics}

In order to derive Eq.~\eqref{FlatBandExpr1}, we perform  a small momentum expansion of tight-binding 
Hamiltonian~\eqref{modelDef}, around the $\Gamma$-point. This yields a continuum model with quadratic dispersions in the normal state
and Fermi wave vectors
\begin{eqnarray}
k^{\alpha}_{\textrm{F}} 
=
- \alpha m \lambda + \sqrt{ ( m \lambda )^2 + 2 m \widetilde{ \mu }} ,
\end{eqnarray}
where $m= - 1 / t $,  $\widetilde{\mu} = \mu - 2 t$, 
and $\alpha \in \{ +, -\}$ labels the two helical Fermi surfaces. 
As in the main text, we consider a (01) edge located at $y=0$, 
where the superconductor and the vacuum occupy the half-spaces $y > 0$ and $y < 0$, respectively.
The zero-energy edge states can be determined by solving the equation $H(k_x,-i\partial_y)\Psi_{k_x}=0$, 
with the wave function ansatz $\Psi_{k_x}=\Psi_{k_x}e^{\kappa y}$. Here, $\mathrm{Re}  [ \kappa  ]$ is the inverse
decay length of the edge state. In the following, we focus on solutions with
positive chirality $\Gamma=+1$, which exist within the interval  $-k_\textrm{F}^+< k_x <-k_\textrm{F}^-$. In that case the secular equation,
$\det \left[ H  (k_x,-i \kappa) \right] = 0$, can be reexpressed as
\begin{align} \label{deteq0}
&\det \left[ D^\dag(k_x,-i\kappa) \right] =
 [ \kappa  k_x \Delta _{\textrm{s}}+ \tilde\mu-\frac{t}{2} (\kappa ^2-k_x^2)  ]^2
 \nonumber\\
& \qquad +\left(\lambda -\kappa  k_x \Delta_\textrm{t}\right)^2\left(\kappa ^2-k_x^2\right) =0 , 
\end{align}
which is a polynomial equation of fourth degree in $\kappa$. 
The nature of the roots of Eq.~\eqref{deteq0} 
can be inferred, to some extent, from the free term 
$a_0=t^2(k_x^2-k_\textrm{F}^-{}^2)(k_x^2-k_\textrm{F}^+{}^2)/4$
of this quartic equation.
An explicit expression for the roots of Eq.~\eqref{deteq0} can be given within the quasi-classical 
approximation.
 For
 $-k_\textrm{F}^+< k_x <-k_\textrm{F}^-$, we have $a_0<0$, 
 and    Eq.~\eqref{deteq0}  for $\Delta_{\textrm{t}} > \Delta_{\textrm{s}}$ has two real roots and two complex conjugate roots.
That is, 
the solutions of Eq.~\eqref{deteq0} are given by\cite{satoPRB2011,Schnyder12} 
\begin{eqnarray} \label{defKappa}
\kappa^+_{\beta}
&=&
- 
k^+_{\perp}  - (-1)^\beta i  \frac{ k^+_{\textrm{F}} }{ k^+_{\perp} }  \sqrt{ \frac{ \widetilde{\Delta}^2_+ [ k_x, i k^+_{\perp} ]    }{ \lambda^2 + 2 \widetilde{\mu} / m }},
\nonumber\\
\kappa^-_{\beta} 
&=&
(-1)^{\beta+1} i k^-_{\perp} - \frac{k^-_{\textrm{F}} }{k^-_\perp} \sqrt{ \frac{ \widetilde{ \Delta }^2_- [ k_x ,  k^-_{\perp} ] }{ \lambda^2 + 2 \widetilde{\mu} / m }} ,\label{Eq:QCroots}
\end{eqnarray}
with $\beta \in \left\{ 1, 2 \right\}$, the transverse momenta $k^+_{\perp} = \sqrt{ k_x^2 - (k_{\textrm{F}}^+)^2 }$ and $k^-_{\perp} = \sqrt{  ( k_{\textrm{F}}^-)^2  - k_x^2}$, and  $\widetilde{\Delta}_{\pm} [ {\bf k} ] = ( k_\textrm{F}^\pm\Delta_\textrm{t} \pm \Delta_\textrm{s} ) k_x k_y$. 
We observe that $\kappa^+_{1}$ and $\kappa^+_{2}$ are purely real, while 
$\kappa^-_{1}$ and $\kappa^-_{2}$  form a complex conjugate pair. 
Furthermore, the maximum decay length $\mathrm{max} \{ - \mathrm{Re} [ \kappa^{\alpha}_\beta ] ^{-1}   \}$
rapidly increases as $k_x \to - k^{\pm}_{\textrm{F}}$. In other words, the wave functions are well confined to the
edge for $k_x$ in the middle of the interval $[ - k^{+}_{\textrm{F}}, - k^{-}_{\textrm{F}} ]$, whereas as $k_x$ approaches the boundaries
of the interval the flat-band states start to penetrate over longer distances  into the bulk.

For each of the four roots $\kappa^{\pm}_{\beta}$, the kernel of the secular equation
is spanned by one basis vector $\psi^{\pm}_{\beta}$, which reads
in the off-diagonal basis, Eq.~\eqref{eq:OffdiagB},
\begin{eqnarray} \label{BasISstates}
&&
\psi^{+}_{\beta}
=
\left(
2 - \beta , 
\frac{ i (2 - \beta) k_{\textrm{F}}^+ }{ k_{\perp}^+ - k_x } , 
\beta -1 , 
 \frac{ i ( \beta -1 ) k_{\textrm{F}^+} }{ k^{+}_{\perp}  - k_x }  
\right)^{\textrm{T}} ,
\nonumber\\
&&
\psi^{-}_{\beta}
=
\left(
1 , 
 \frac{ - k_{\textrm{F}}^- }{ i k_x + (-1)^{\beta} k^-_{\perp} },
0 ,
0   
\right)^{\textrm{T}} ,
\end{eqnarray}
with $\beta \in  \{ 1, 2\}$. With this, the ansatz for the flat-band edge states
can be written as a linear combination of the basis states~\eqref{BasISstates}
\begin{align}
\Psi_{k_x} =
\sum_{\alpha \in \{ +, - \} }
\sum_{\beta \in \{ 1, 2\} }
C^\alpha_\beta \psi^{\alpha}_{\beta} e^{  \kappa^{\alpha}_{\beta} y }  e^{i k_x x}\label{eq:WaveFunAnsatz}
\end{align}
where the coefficients $C^\alpha_\beta$
are fixed by the boundary conditions
\begin{align} \label{Bcond}
\Psi (k_x, y=0) = 0, \ \ \Psi (k_x, y=\infty) = 0.
\end{align} 
The latter condition implies $C^+_2=0$, since $\textrm{Re}[\kappa^+_2]>0$,
but there exists a nonzero solution for $(C^+_1, C^-_1, C^-_2)$ that satisfies the boundary conditions. 
After some algebra, we find that in the off-diagonal basis, Eq.~\eqref{eq:OffdiagB}, the  zero-energy edge-state wave function with
positive chirality $\Gamma = +1$ is given by
$\Psi^+_{k_x} = \left( \chi_{k_x}   , 0 \right)^{\textrm{T}}$,
with
\begin{equation} \label{EdgwWaveFunA}
\chi_{k_x}  
=
\begin{pmatrix}
- 2 k^+_{\textrm{F}} k^-_{\perp} e^{\kappa^+_1 y}  + b_1 e^{\kappa^-_2 y} + b_1^{\ast} e^{\kappa^-_1 y}   \cr
 2 i a_1  k^-_\perp e^{\kappa^+_1 y}    - a^{\ }_2 i e^{\kappa_1^- y}  - a^{\ast}_2 i e^{\kappa_2^- y} \cr 
\end{pmatrix} e^{i k_x x} ,
\end{equation}
where $a_1 = k^+_{\perp} + k_x$, $a_2 = a_1 ( k^-_{\perp} + i k_x ) + i k^{+}_{\textrm{F}} k^-_{\textrm{F}}$, and
$b_1 = k^+_{\textrm{F}} ( k^-_{\perp} + i k_x ) + i k^-_{\textrm{F}} a_1 $.

Similarly, we can derive solutions of the equation $H(k_x,-i\partial_y)\Psi_{k_x}=0$ with negative chirality $\Gamma=-1$, which 
exist within the interval $k^+_{\textrm{F}} <k_x< k^-_{\textrm{F}}$. Repeating similar steps as above,
we find that the negative chirality edge-state wave function is given by
$\Psi^-_{k_x}  = \left( 0, \eta_{k_x} \right)^{\textrm{T}}$, with
\begin{equation} \label{EdgwWaveFunB}
\eta_{k_x} 
=
\begin{pmatrix}
- 2 i \widetilde{a}_1 k^-_{\perp}  e^{\kappa^+_1 y} + \widetilde{a}_2 i e^{\kappa^-_1 y}  + \widetilde{a}^{\ast}_2 i e^{\kappa^-_2 y}  \cr
 2 k^+_{\textrm{F}}  k^-_{\perp} e^{\kappa^+_1 y}  - \widetilde{b}_1^{\ast} e^{\kappa^-_1 y} - \widetilde{b}^{\ }_1e^{\kappa^-_2 y} \cr
\end{pmatrix} e^{i k_x x}  ,
\end{equation}
where $\widetilde{a}_1 = k^+_{\perp} - k_x $, $\widetilde{a}_2 =  \widetilde{a}_1 ( k^-_{\perp} - i  k_x  ) + i k^-_{\textrm{F}} k^+_{\textrm{F}}    $, and 
$\widetilde{b}_1 = k^+_{\textrm{F}} ( k^-_{\perp} - i  k_x   ) + i  k^-_{\textrm{F}} \widetilde{a}_1$. 
As expected,  the wave functions $\Psi^+_{k_x}=(\chi_{k_x},0)$ and $\Psi^-_{k_x}=(0,\eta_{k_x}) $ transform into each other by time-reversal symmetry, i.e.,
\begin{eqnarray}
\begin{pmatrix}
0 & i \sigma_y \cr
i \sigma_y & 0 \cr
\end{pmatrix} \Psi^{+}_{k_x} (y) 
=
\left[ \Psi^{-}_{- k_x} (y) \right]^{\ast}  ,
\end{eqnarray}
where we have used the fact that $\kappa^+_1$ is purely real and $( \kappa^-_{1}, \kappa^-_{2})$ are complex conjugate partners. 

Using Eqs.~\eqref{EdgwWaveFunA} and~\eqref{EdgwWaveFunB}, the expectation value of the spin operator  $\widetilde{S}^{\mu}$ 
can be computed in a straightforward manner. 
In particular, we find that 
$
\left\langle \Psi^+_{k_x}  \right|   \widetilde{S}^{x} \left| \Psi^+_{k_x}  \right\rangle=0,
$
since the first component of $\chi_{k_x}$, Eq.~\eqref{EdgwWaveFunA}, is purely real, whereas
the second component of $\chi_{k_x}$ is purely imaginary. 
Likewise, we have $
\left\langle \Psi^-_{k_x}  \right|   \widetilde{S}^{x} \left| \Psi^-_{k_x}  \right\rangle=0
$.

In closing, we remark that for other edge orientations the flat-band states can be derived in a similar manner as above. 
For any given edge orientation we can define the momentum parallel to the edge as
$k_\parallel=k_x\sin(\varphi)+k_y\cos(\varphi)$. Correspondingly, the secular equation for
the positive chirality edge states reads
$\det [ D^\dag(k_\parallel,-i\kappa)]=0$. 
Within the continuum approximation, we find that for dominant triplet pairing, $\Delta_{\textrm{t}} > \Delta_{\textrm{s}}$, 
the flat bands are always of single degeneracy.\cite{Brydon10,Schnyder12} 
Zero-energy states appear in regions of the edge Brillouin zone that are bounded
by the projected nodal points.  
Hence, the density of edge states per unit length can be approximated by
\begin{eqnarray} 
\frac{ d N}{d l} 
=
 \left|  k_0^-  \sin \varphi -  k_0^+ \cos \varphi \right|
- \left| k_0^- \cos \varphi - k_0^+ \sin \varphi  \right| , \hspace{0.9cm} 
\end{eqnarray}
where $\varphi\in[0,\pi/4[$ is the angle between 
the considered edge orientation and the nearest (01) or (10) direction.

\section{Recursive Green's function technique}
\label{appA}

\begin{figure}[t!]
\centering
\includegraphics[clip,angle=0,width=1\columnwidth,bb = 0cm 2cm 36cm 24cm]{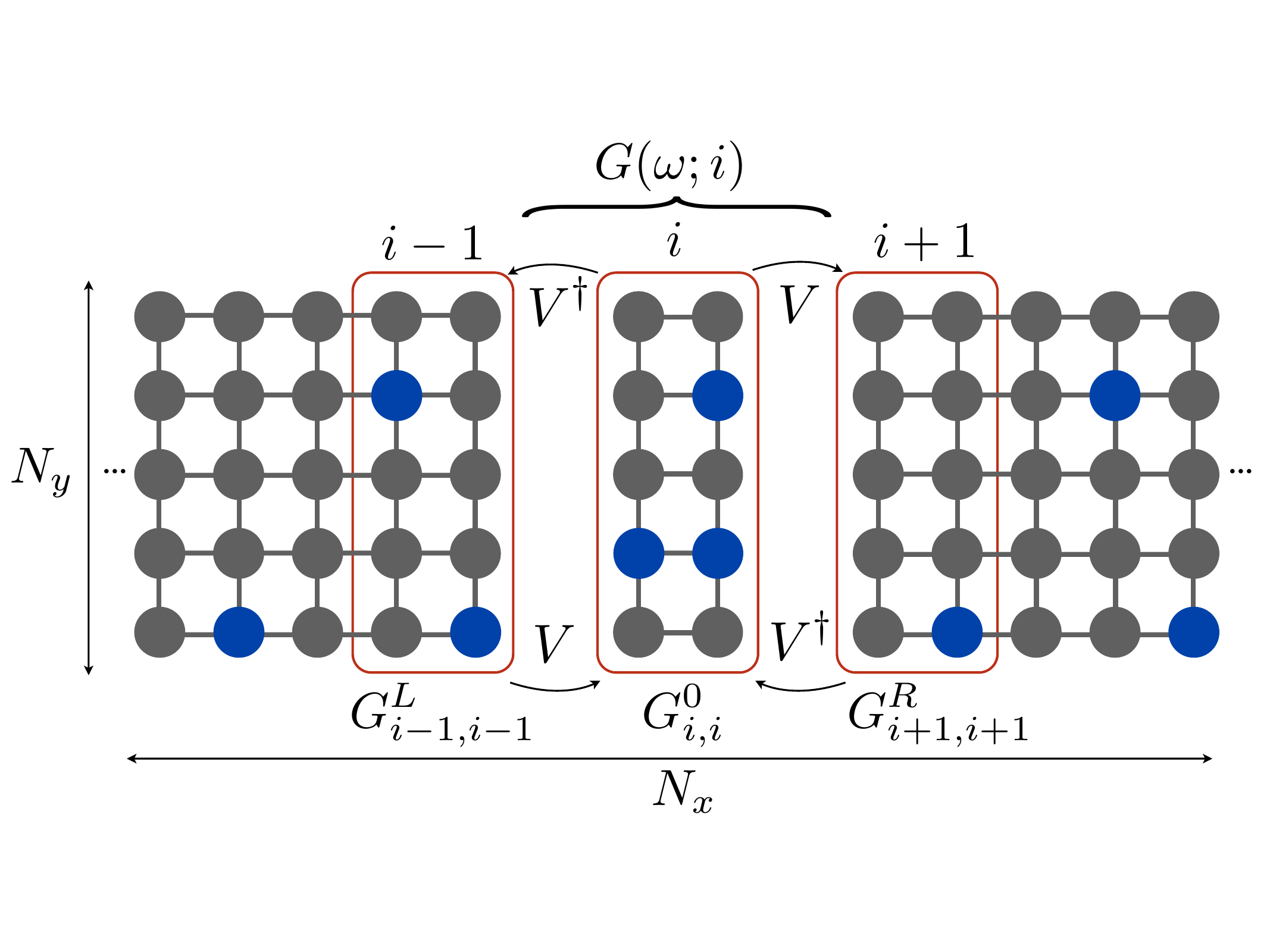}
\caption{\label{fig6L}  
(Color online) 
 Scheme of the recursive Green's function method. To accommodate second neighbor hopping an effective building block of two columns is considered, as well as an effective coupling $V$. The local Green's function $G(\omega;i)$ is calculated by attaching to the block $i$ the left ($G^L$) and right ($G^R$) ribbons according to Eq.~(\ref{recrelation}). The blue sites represent random on-site disorder potentials, defined by $H_\mathrm{imp}^\beta(x)$.}  
\end{figure}

The local density of states of a disordered superconductor can be efficiently computed 
using the recursive Green's function technique.\citep{PotterLee11,LeeFisher81} This is achieved by considering a discrete ribbon in real space with width of $\mathcal{N}_y$ sites and length of $\mathcal{N}_x$ sites. Let us define the block Hamiltonian $H^n$ corresponding to the coupling of two columns spaced by $n$ lattice sites in the $x$ direction,
\begin{align}
H^n=\frac{1}{(2\pi)^2}\int d^2\mathbf{k}H_\mathbf{k}e^{ik_y(y-y')}e^{ink_x},
\end{align}
where $H^0$ corresponds to a free column Hamiltonian, while $H^1$ and $H^2$ are the coupling to the nearest and next nearest neighbouring columns, respectively.
We have considered in our simulations on-site impurities as defined in Eq.~\eqref{EqImpurePo}. For each ribbon's column, i.e., fixed $x$ in ${\bf r}=(x,y)$, we have
\begin{align}
H_\mathrm{imp}^\beta(x)=\sum_{i=1,N_\mathrm{imp}}\Phi^\dag_{\bf r}v_\mathrm{imp}(\mathbf{r_i})\delta_{\mathbf{r},\mathbf{r}_i}S^\beta\Phi_{\bf r},
\end{align}
with impurity positions, ${\bf r_i}$, defined according to the disorder distribution, see Sec. \ref{sec:dis}. To accommodate next-nearest neighbour hopping in a convenient way, we redefine the building blocks of the ribbon to be $8\mathcal{N}_y\times 8\mathcal{N}_y$-matrices incorporating two columns as the building blocks, by writing
\begin{align}
H^\beta_i&=\begin{pmatrix}
H^0+H_\mathrm{imp}^\beta(2i-1)
&H^1\\H^1{}^\dag&H^0+H^\beta_\mathrm{imp}(2i)
\end{pmatrix},\nonumber\\
V&=\begin{pmatrix}
H^2&0\\ H^1&H^2
\end{pmatrix}.
\end{align}
Here, the block index $i$ runs from 1 to $\mathcal{N}_x/2$, i.e., with spacing of 2 lattice sites. Dyson's equation for $G(\omega;i)\equiv G_{i,i}$ takes the form
\begin{eqnarray}
G_{i,i}=G^{0}_{i,i}+G^{0}_{i,i}VG^L_{i-1,i}+G^{0}_{i,i}V^\dag G^R_{i,i+1},
\end{eqnarray}
where the $\omega$ dependence is suppressed for simplicity, $G^0_{i,i}=[\omega+i\eta-H_i^\beta]^{-1}$ stands for the uncoupled block at $i$, while $G_{i,j}^{L/R}$ represent the ribbons to its left and right,
\begin{eqnarray}
G^L_{i-1,i}&=G^L_{i-1,i-1}VG_{i,i}, \hspace{0.4cm}
G^R_{i,i+1}&=G^R_{i+1,i+1}V^\dag G_{i,i}.\hspace{0.5cm}
\end{eqnarray}

\noindent The side ribbons are calculated recursively, using the relations
\begin{align}
G^{L(n)}_{n,n}&=\left[\omega+i\eta-H_{n-1}^\beta-V^\dag G^{L(n-1)}_{n-1,n-1}V\right]^{-1},\nonumber\\
G^{R(n)}_{n,n}&=\left[\omega+i\eta-H_{n+1}^\beta-V G^{R(n+1)}_{n+1,n+1}V^\dag\right]^{-1}, 
\end{align}
at iteration step $n$, where $n\in\{1,...,i-1\}$ and $n\in\{\mathcal{N}_x/2,...,i+1\}$ for $G^L$ and $G^R$, respectively.
Finally, the local Green's function can be reduced to
\begin{align}
G_{i,i}=\left[\omega+i\eta-H_i^\beta-\Sigma^L_i-\Sigma^R_i\right]^{-1} \label{recrelation},
\end{align}
with self-energies, $\Sigma^{L/R}_i$, given by
\begin{align}
\Sigma^{L}_i=V^\dag G^L_{i-1,i-1}V,&&
\Sigma^{R}_i=V G^R_{i+1,i+1}V^\dag.
\end{align}
The local density of states can then be easily computed by Eq.~\eqref{LDOSrgf}, where $G(\omega;x,y)$ are obtained from the diagonal entries of $G(\omega;i)$.

\bibliography{robustrefsV3}

\begin{thebibliography}{58}
\bibitem[{foo({\natexlab{c}})}]{footnote2}
\bibinfo{note}{For each member of the ensemble of irregularly shaped dots the
  weights $w_i$ and phases $\phi_i$ in Eq.~\eqref{DotShapes} are randomly
  chosen from the intervals $]0,0.05[$ and $]0,2\pi[$, respectively.}
\end{thebibliography}

 \end{document}